\documentstyle [12pt]{article}
\def\int {\intop \limits}

\def\fnote#1{\footnote}
\begin{document}
\newcommand{\dst}[1]{\displaystyle{#1}}
\newcommand{\barl}{\begin{array}{rl}}
\newcommand{\ball}{\begin{array}{ll}}
\newcommand{\ear}{\end{array}}
\newcommand{\barc}{\begin{array}{c}}
\newcommand{\sne}[1]{\displaystyle{\sum _{#1} }}
\newcommand{\sn}[1]{\displaystyle{\sum ^{\infty }_{#1} }}
\newcommand{\ini}[1]{\displaystyle{\int ^{\infty }_{#1}}}
\newcommand{\myi}[2]{\displaystyle{\int ^{#1}_{#2}}}
\newcommand{\inn}{\displaystyle{\int }}
\newcommand{\be}{\begin{equation}}
\newcommand{\ee}{\end{equation}}
\newcommand{\aq}[1]{\label{#1}}

\vspace*{4.0cm}
\centerline{\Large {\bf Quantum theory of transition radiation}}
\vskip .25cm
\centerline{\Large {\bf and transition pair creation}}
\vskip .5cm
\centerline{\large{\bf V. N. Baier and V. M. Katkov}}
\centerline{Budker Institute of Nuclear Physics,
 630090 Novosibirsk, Russia}
\vskip 2.0cm
\begin{abstract}
Theory of the transition radiation and the transition pair creation
is developed in the frame of QED. The spectral-angular distributions of
probability of the transition
radiation and of the transition pair creation are found. The total energy
losses of and the total probability of pair creation are calculated and
analyzed. Features of radiation and pair creation processes in a
superdence medium (typical for white dwarfs) are discussed.

\end{abstract}

\newpage

1.~The transition radiation arises at
uniform and rectilinear motion of a charged
particle when it intersects a boundary of two different media (in general
case, when it moving in a nonuniform medium or near such medium).
This phenomenon \cite{1} was actively investigated
during a few last decades (see, e.g. reviews \cite{2}, \cite{3}) and
widely used in transition radiation detectors.
The existing theory of the transition radiation is based on
the classical electrodynamics.
The quantum theory of the transition radiation is of evident interest.
An analysis in the frame of quantum electrodynamics indicates existence of
the crossing process: electron-positron pair creation by a photon on a
boundary of two different media. We shall call this process as
the transition pair creation.

It turns out that the quasiclassical operator method developed by authors
is adequate for consideration of the transition radiation. The probability
of the process has a form (see \cite{4}, p.63, Eq.(2.27))
\begin{equation}
dw=\frac{e^2}{(2\pi)^2} \frac{d^3k}{\hbar \omega}
\int_{-\infty}^{\infty}dt_2\int_{-\infty}^{\infty}dt_1 R^{\ast}(t_2) R(t_1)
\exp\left[-\frac{i}{\varepsilon'}\int_{t_1}^{t_2}
\left(kp-\frac{\hbar k^2}{2} \right)dt \right],
\label{1}
\end{equation}
where $p=p^{\mu}=(\varepsilon, {\bf p})$ is the 4-momentum of the initial
electron, $k=k^{\mu}=(\omega, {\bf k})$ is the 4-momentum of the radiated
photon, in a medium $|{\bf k}|=n \omega$, $n$ is the refractive index,
$\varepsilon'=\varepsilon-\hbar \omega$, the matrix element $R(t)$
is defined by the structure of a current, we employ units such that
$c=1$. Here we took into account
the term $k^2$ in the exponent (this is result of the disentanglement
of the operator expression, see Eq.(2.23) of \cite{4}) which is essential in
the considered case since we consider radiation in a medium (this term was
skipped in our paper \cite{5}) and we use
the representation Eq.(2.24) of \cite{4} because we are dealing with
nonuniform case. For electrons (spin 1/2 particle) one has
\begin{eqnarray}
&& R(t)=\frac{m}{\sqrt{\varepsilon \varepsilon'}}
\overline{u}_{s_{f}}({\bf p'})\hat{e}^{\ast}u_{s_{i}}({\bf p})
=\varphi_{s_{f}}^{+}\left(A(t)+i\mbox{\boldmath$\sigma$}{\bf B}(t) \right)
\varphi_{s_{i}}, \nonumber \\
&& A(t)=\frac{{\bf e}^{\ast}{\bf p}(t)}{2\sqrt{\varepsilon \varepsilon'}}
\left[\sqrt{\frac{\varepsilon'+m}{\varepsilon+m}}+
\sqrt{\frac{\varepsilon+m}{\varepsilon'+m}} \right]
\simeq \frac{1}{2}\left(1+\frac{\varepsilon}{\varepsilon'} \right)
{\bf e}^{\ast}\mbox{\boldmath$\vartheta$}, \nonumber \\
&& {\bf B(t)}=\frac{1}{2\sqrt{\varepsilon \varepsilon'}}
\left[\sqrt{\frac{\varepsilon'+m}{\varepsilon+m}} \left({\bf e}^{\ast}
\times {\bf p}(t) \right)
- \sqrt{\frac{\varepsilon+m}{\varepsilon'+m}} \left({\bf e}^{\ast}
\times ({\bf p}(t)-\hbar {\bf k}) \right)\right] \nonumber \\
&&\simeq \frac{\hbar \omega}{2\varepsilon'}\left({\bf e}^{\ast} \times
\left(\frac{{\bf n}}{\gamma}- \mbox{\boldmath$\vartheta$}\right) \right),
\label{2}
\end{eqnarray}
here ${\bf e}$ is the vector of polarization of photon (Coulomb gauge is
used), four-component spinors $u_{s_f}, u_{s_i}$ and
two-component spinors $\varphi_{s_f}, \varphi_{s_i}$ describe
the initial ($s_i$) and final ($s_f$) polarization of the electron,
$\mbox{\boldmath$\vartheta$} =
v^{-1}\left({\bf v}-{\bf n}({\bf n}{\bf v}) \right) \simeq {\bf v}_{\perp}$,
${\bf v}_{\perp}$ is the component of particle velocity transverse to the
vector ${\bf n}={\bf k}/|{\bf k}|$,
$\gamma=\varepsilon/m$ is the Lorentz factor.
The final expressions in (\ref{2}) are given for
radiation of ultrarelativistic electrons, they are written down
with relativistic accuracy (terms $\sim 1/\gamma$ are neglected) and in
the small angle approximation. For the rectilinear motion radiation
arises because of variation of the refractive index $n(\omega)$,
e.g. at intersection
of a boundary of two different media. The main contribution gives a region
of high frequencies where
\begin{equation}
n(\omega)=1-\frac{\omega_0^2}{2 \omega^2},\quad
\omega_0^2=\frac{4\pi e^2 N}{m},
\label{3}
\end{equation}
where $N$ is the density of electrons in a medium, $\omega_0$ is the plasma
frequency.

2.~Here we consider the transition radiation in forward direction
at normal incidence of the relativistic particle on the boundary between
vacuum and a medium. In this case the photon mass squared can be written as
\begin{equation}
\hbar^2 k^2=(\hbar \omega_0)^2 g(t),
\label{4}
\end{equation}
where the function $g(t)$ describes variation of the density of a medium
on the projectile trajectory.
The combination $R^{\ast}(t_2) R(t_1)$ in (\ref{1})
can be presented in a form
\begin{equation}
R^{\ast}(t_2) R(t_1)=\frac{1}{4}{\rm Tr}\left[(1+\mbox{\boldmath$\sigma$}
\mbox{\boldmath$\zeta$}_i)\left(A^{\ast}(t_2)-
i\mbox{\boldmath$\sigma$}{\bf B^{\ast}}(t_2) \right)
(1+\mbox{\boldmath$\sigma$}
\mbox{\boldmath$\zeta$}_f)
\left(A(t_1)+i\mbox{\boldmath$\sigma$}{\bf B}(t_1) \right)\right],
\label{5}
\end{equation}
where $(1+\mbox{\boldmath$\sigma$}\mbox{\boldmath$\zeta$})$ is two-dimensional
polarization density matrix, we neglect here
change of the electron spin during
radiation process. If we are not interested in the initial and final electron
polarizations, then
\begin{equation}
\frac{1}{2}\sum_{s_i,s_f}^{}R^{\ast}(t_2) R(t_1)= A^{\ast}(t_2)A(t_1)+
{\bf B}^{\ast}(t_2){\bf B}(t_1).
\label{6}
\end{equation}
Summing over the photon polarizations $\lambda$ we have
\begin{equation}
\sum_{\lambda=1}^{2}\left( A^{\ast}(t_2)A(t_1)+
{\bf B}^{\ast}(t_2){\bf B}(t_1)\right)=\frac{1}{2\varepsilon'^2}\left[
\frac{(\hbar \omega)^2}{\gamma^2}+(\varepsilon^2+\varepsilon'^2)
\mbox{\boldmath$\vartheta$}(t_2)\mbox{\boldmath$\vartheta$}(t_1) \right].
\label{7}
\end{equation}
This form of $R^{\ast}(t_2) R(t_1)$ was used in \cite{6}.
For the rectilinear trajectory
\begin{equation}
\sum_{\lambda=1}^{2}(A^{\ast}(t_2)A(t_1)+
{\bf B}^{\ast}(t_2){\bf B}(t_1))=\frac{1}{2\varepsilon'^2}\left[
r_1+r_2\vartheta^2 \gamma^2 \right],
\label{8}
\end{equation}
where $\vartheta$ is the angle between vectors ${\bf p}$ and ${\bf k}$,
\[
r_1=\frac{(\hbar \omega)^2}{\varepsilon^2},\quad r_2=1+
\frac{\varepsilon'^2}{\varepsilon^2}.
\]
In the case considered one can expand
\begin{eqnarray}
&& kp = \omega \varepsilon (1-{\bf nv}) \simeq \omega \varepsilon
\left(\frac{1}{2\gamma^2}+
\frac{\vartheta^2}{2}+\frac{k^2}{2\omega^2} \right), \nonumber \\
&& \hbar kp-\frac{(\hbar k)^2}{2} \simeq
\frac{\hbar \omega \varepsilon}{2\gamma^2}
\left[1+\gamma^2 \vartheta^2+\frac{k^2}{m^2}
\frac{\varepsilon(\varepsilon-\hbar \omega)}{\omega^2} \right].
\label{9}
\end{eqnarray}
Substituting the results obtained ((\ref{6}), (\ref{8}) and (\ref{9})) into
Eq.(\ref{1}) we obtain the spectral-angular distribution of
the probability of the transition radiation
\begin{equation}
\frac{dw}{d\hbar \omega dy}=\frac{e^2}{2\pi \hbar^2 \omega}
\left(r_1+r_2 y \right) |M(y)|^2,
\label{10}
\end{equation}
where
\begin{equation}
M(y)=\int_{-\infty}^{\infty} \exp\left[-i\int_{0}^{x}(1+y+
\varphi(t))dt \right] dx,\quad \varphi(t)=\frac{\omega_0^2}{\omega^2}
\frac{\varepsilon \varepsilon'}{m^2} g(t) \equiv \varphi_0 g(t).
\label{11}
\end{equation}
Here we introduced the angular variable $y=\gamma^2 \vartheta^2$
and substitution of the variable
$\displaystyle{\frac{\omega m^2}{2\varepsilon \varepsilon'}t \rightarrow t}$
in the expression for $M(y)$ is performed.

An important case is the transition radiation on the boundary between
vacuum and the medium. In this case $g(t) \rightarrow \vartheta(t)$ and
we take integral over angle
\begin{eqnarray}
&& M(y)=i\left(\frac{1}{1+y}-\frac{1}{\kappa+y} \right),\quad
\kappa=1+\varphi_0, \nonumber \\
&& \frac{dw_{tr}}{d\hbar \omega}=\frac{e^2}{2\pi \hbar \omega}
\left\{r_1 \left[1+\frac{1}{\kappa}-\frac{2}{\kappa-1}\ln \kappa \right]+
r_2\left[\left(1+\frac{2}{\kappa-1} \right)\ln \kappa -2 \right] \right \}.
\label{12}
\end{eqnarray}
The results obtained ((\ref{10})-(\ref{12})) is the quantum generalization
of the theory of the transition radiation. In the classical limit
$\hbar \omega \ll
\varepsilon$ one has $r_1 \rightarrow  0,~r_2 \rightarrow 2$,
\newline $\displaystyle{\varphi_0 \rightarrow \frac{\omega_0^2}
{\omega^2}\gamma^2 =\kappa_0^2}$
and we have from (\ref{12}) the known expression for the spectral
distribution of radiated
energy of the transition radiation on the boundary between
vacuum and the medium $dE/d\omega \equiv \hbar \omega~dw_{tr}/d\omega$
in the classical theory (see e.g.
\newline \cite{2}-\cite{3}). The spectrum (\ref{12})
for the case $\hbar \omega_0=2m,~\varepsilon=100~MeV$ is shown in Fig.1
(curve 1), for comparison the classical spectrum (curve 2) is given.
The classical probability is always larger than the exact one and
remains finite at $\hbar \omega=\varepsilon$.

Process of the transition radiation in frame of the quantum theory
was considered many years ago in \cite{7}. The spectral distribution
of the probability of the transition radiation $w(\omega)$
obtained in cited paper differs from Eq.(\ref{12}).
It should be noted that Eq.(14) of \cite{7} for $w(\omega)$
doesn't satisfy the symmetry relation with respect permutation
$\varepsilon \leftrightarrow \varepsilon'$ in the crossing channel
(after application the substitution law, see (\ref{17}) below).

When $\displaystyle{\varphi_0=\frac{\omega_0^2}{\omega^2}
\frac{\varepsilon \varepsilon'}{m^2} \ll 1}$ the spectral energy losses are
\begin{equation}
\frac{dE}{d\omega}=\frac{e^2}{12\pi}\varphi_0^2 \left[2r_1+r_2 \right]=
= \frac{e^2}{12\pi} \frac{\omega_0^4}{m^4}
\frac{\varepsilon'^2(2(\hbar \omega)^2+\varepsilon^2+
\varepsilon'^2)}{\omega^4}.
\label{13}
\end{equation}
So for $\hbar \omega_0 \ll m$ the hard part of the spectrum
of the transition radiation (for $\hbar \omega \sim \varepsilon$)
is suppressed as $(\hbar \omega_0)^4/m^4$ (for the case considered
we have power suppression)
tending to zero at the end of the spectrum
as $\varepsilon'^2$.

Integrating (\ref{12}) over photon energies we obtain after rather cumbersome
calculation the total energy losses $\Delta E$ when the electron intersects
the boundary between vacuum and the medium
\begin{eqnarray}
&& \Delta E = \int_{0}^{\varepsilon} \hbar \omega \frac{dw_{tr}}{d\hbar \omega}
d\hbar \omega \nonumber \\
&& = \frac{e^2}{\pi \hbar }\varepsilon \left\{2a+
\frac{4a}{3}(1-2a)\left[-2+4a \ln 4a+ (1+(1-2a)4a)J(a) \right] \right\},
\label{14}
\end{eqnarray}
where $\displaystyle{a=\frac{(\hbar \omega_0)^2}{4m^2}
=\frac{\hbar^2 k^2}{4m^2}}$,
\[ J(a) = \left\{
\begin{array}{lr}
 \displaystyle{ \frac{1}{\sqrt{a(1-a)}}
\left(\frac{\pi}{2}-\arctan \sqrt{\frac{a}{1-a}}~ \right)},~&  a<1; \\
 1,                                                        ~&  a=1; \\
 \displaystyle{ \frac{1}{\sqrt{a(a-1)}}
 \ln \left(\sqrt{a}+\sqrt{a-1)} \right)},                  ~&   a>1.
\end{array} \right.
\]

In the limit $a \ll 1$ we have from (\ref{14})
\begin{equation}
\Delta E = \frac{e^2}{3}\omega_0 \gamma\left(1-\frac{3}{2\pi}
\frac{\hbar \omega_0}{m} \right ),
\label{14a}
\end{equation}
here the first term is the known classical result and the second term is
the first quantum correction.
In the opposite limit $a \gg 1$ we find from (\ref{14})
for the total energy losses $\Delta E$
\begin{equation}
\Delta E = \frac{2 e^2}{3\pi}\varepsilon \left(\ln 4a -\frac{1}{6} \right).
\label{14b}
\end{equation}
The total energy losses $\Delta E$ when the electron intersects
the boundary between vacuum and the medium are shown in Fig.2 vs
$\displaystyle{a=\frac{(\hbar \omega_0)^2}{4m^2}}$.

For any matter
on the Earth $\hbar \omega_0 < 100~eV$ and it follows from above that
photons of the
transition radiation are soft ($\hbar \omega_{tr}
\ll \varepsilon$).
However for the matter with density $\varrho \sim 10^8 g/cm^3$ (white dwarfs)
one has $\hbar \omega_0 \sim m$ and the region of photon energies
$\hbar \omega_{tr} \sim \varepsilon$ is not suppressed.
So the quantum theory of the
transition radiation may have astrophysical applications.

The polarization of the transition radiation can be found from (\ref{6}).
We introduce two polarization vectors
\begin{equation}
{\bf e}_1=\frac{{\bf n}\times({\bf s}\times{\bf n})}
{|{\bf n}\times({\bf s}\times{\bf n})|}, \quad
{\bf e}_2=\frac{{\bf s}\times{\bf n}}
{|{\bf s}\times{\bf n}|},
\label{15}
\end{equation}
where ${\bf s}={\bf v}/|{\bf v}|$. Substituting the amplitudes $A(t)$ and
${\bf B}(t)$ from Eq.(\ref{2}) one obtains the Stokes's parameters
\begin{equation}
\xi_1=\xi_2=0,\quad \xi_3= \frac{2\varepsilon \varepsilon' \vartheta^2}
{(\hbar \omega)^2/\gamma^2+
\left(\varepsilon^2+\varepsilon'^2 \right)\vartheta^2}.
\label{16}
\end{equation}
In the classical limit one has $\hbar \omega \rightarrow 0,~\varepsilon'
\rightarrow \varepsilon$  and we arrive to the
known result \cite{2} that the transition radiation
is completely linearly polarized ($\xi_3 = 1$) in the radiation plane.

3.~The crossing process for the transition radiation is the transition
pair creation: when a photon intersects the boundary of two different media
(in the general case, when it is moving in a nonuniform
medium) the photon mass squared
$(\hbar k)^2 \neq 0$ changes and creation of the
electron-positron pair becomes
possible. The probability of the pair creation can be obtained from the
probability of radiation with help of the substitution law:
\begin{equation}
d^3k \rightarrow d^3p/\hbar^3,\quad \omega \rightarrow -\omega,
\quad \varepsilon
\rightarrow -\varepsilon
\label{17}
\end{equation}
Starting from (\ref{10}) and (\ref{11}) we have for
the spectral-angular distribution of
probability of the transition pair creation for the created electron
\begin{equation}
\frac{dw}{d\varepsilon dy}=\frac{e^2}{2\pi \hbar^2 \omega}
\left(s_1+s_2 y \right) |M(y)|^2,
\label{18}
\end{equation}
where
\begin{equation}
M(y)=\int_{-\infty}^{\infty} \exp\left[-i\int_{0}^{x}(1+y-
\varphi(t))dt \right] dx,\quad s_1=1,
~s_2=\frac{\varepsilon^2+\varepsilon'^2}{(\hbar \omega)^2},
~\varepsilon+\varepsilon'=\hbar \omega,
\label{19}
\end{equation}
here $\varepsilon~(\varepsilon')$ is the energy of the created electron
(positron), $y=(\gamma \vartheta)^2$, $\vartheta$ is the angle between
momentum of the initial photon and the momentum of the created electron,
$\varphi (t)$ is defined in (\ref{11}).

An important case is the transition pair creation on the boundary between
vacuum and the medium. In this case $g(t) \rightarrow \vartheta(t)$ and
we take integral over angle
\begin{eqnarray}
&& M(y)=i\left(\frac{1}{1+y}-\frac{1}{\chi+y} \right),\quad
\chi=1-\varphi_0,\quad \varphi_0= \frac{\omega_0^2}{\omega^2}
\frac{\varepsilon \varepsilon'}{m^2}\nonumber \\
&& \frac{dw_{p}}{d\varepsilon}=\frac{e^2}{2\pi \hbar^2 \omega}
\left\{s_1 \left[1+\frac{1}{\chi}-\frac{2}{\chi-1}\ln \chi \right]+
s_2\left[\left(1+\frac{2}{\chi-1} \right)\ln \chi -2 \right] \right \}.
\label{20}
\end{eqnarray}
As one can expect the probability (\ref{20}) is symmetrical with respect
energies $\varepsilon$ and $\varepsilon'$.

Integrating (\ref{20}) over the electron energy we obtain
the total probability of the transition pair creation
\begin{equation}
w_p=\int_{0}^{\omega} \frac{dw_{p}}{d\varepsilon} d\varepsilon =
\frac{e^2}{2\pi \hbar }\left[\frac{5+2a-4a^2}{3a(1-a)}\sqrt{\frac{1-a}{a}}\arctan
\sqrt{\frac{a}{1-a}}-\frac{5}{3a}-\frac{16}{9}  \right],
\label{21}
\end{equation}
where  $a<1,~a$ is defined in Eq.(\ref{14}).

In the limit $a \ll 1$ one has from (\ref{21})
\begin{equation}
w_p=
\frac{e^2}{2\pi \hbar }\left[\frac{8}{35}a^2+\frac{256}{945}a^3+
O(a^4) \right].
\label{22}
\end{equation}
This means that in this limit the pair creation probability is damped
$\propto (\hbar \omega_0/m)^4$~(this is the only result obtained
in \cite{7} for pair creation, however with wrong coefficient).

The total probability of the transition pair creation $w_p$ is shown in Fig.3
as the function of $a=(\hbar \omega_0)^2/4m^2=(\hbar k)^2/4m^2$.
It is seen that $w_p$ grows very fast with $a$ increase.

At $a \rightarrow 1$ Eq.(\ref{21})
can be written as
\begin{equation}
w_p \simeq
\frac{e^2}{4 \hbar}\left[\frac{1}{\sqrt{1-a}}-\frac{80}{9\pi}\right].
\label{23}
\end{equation}
At $a=1$ the value of $w_p$ becomes divergent. We have to recall
that $w_p$ is the total probability of pair creation
for the infinite time. If takes into account an absorption
(or the imaginary part
of the refraction index of the medium) the value of $w_p$ will be finite.
\newline At $a > 1 ((\hbar k)^2 > 4m^2)$ the photon
becomes unstable since the channel
of decay $\gamma \rightarrow e^+e^-$pair will be open and just this process
gives the contribution into the absorption. The corresponding expression
for the probability the process per unit time one can obtain using
Eqs.(\ref{18})-(\ref{19}):
\begin{eqnarray}
&& M(y)=2\pi \delta\left(1+y-
a\frac{4\varepsilon \varepsilon'}{(\hbar \omega)^2} \right)
\nonumber \\
&& \frac{dw_{p}}{dt d\varepsilon}=\frac{e^2}{\hbar \omega}
\frac{m^2\hbar \omega}{2\varepsilon \varepsilon'}
\int_{0}^{\infty}
\left(1+y \frac{\varepsilon^2+\varepsilon'^2}{(\hbar \omega)^2} \right)
\delta\left(1+y-
a\frac{4\varepsilon \varepsilon'}{(\hbar \omega)^2} \right) dy.
\label{24}
\end{eqnarray}
Here we returned to the standard time making the inverse substitution
$\displaystyle{t \rightarrow \frac{\omega m^2}{2\varepsilon \varepsilon'}t}$
(see Eq.(\ref{11})). This expression can be obtained also if one considers
large but finite part of the projectile trajectory in the medium. In this case
the pole term (see (\ref{20})) dominates
\begin{eqnarray}
&& \hspace{-15mm} M(y) \simeq \frac{i}{\chi+y}\left[\exp (i(\chi+y)T)-1 \right],\quad
\int_{0}^{\infty}(s_1+s_2y)|M(y)|^2 dy 
\nonumber \\
&& \hspace{-15mm} \simeq \int_{0}^{\infty}(s_1+s_2y)
\frac{1-\cos (\chi+y)T}{(\chi+y)^2}
\simeq (s_1-s_2\chi)2\pi T;~
\chi <0,~ -\chi T =-\chi \frac{m^2\omega}{2\varepsilon \varepsilon'}t
\gg 1.
\label{24a}
\end{eqnarray}
The same result will be found after integration over $y$ in (\ref{24}).

 Introducing the variable $x=\varepsilon/(\hbar \omega)$
and passing to the variable $z=2x-1$ we find
\begin{eqnarray}
\nonumber \\
&& \frac{dw_{p}}{dt}=\frac{e^2 m^2}{\hbar \omega}
\int_{0}^{\beta}
\left\{2+(1+z^2)\left[a(1-z^2)-1 \right] \right\}\frac{dz}{1-z^2} \nonumber \\
&& =\frac{e^2 m^2}{\hbar \omega}
\int_{0}^{\beta}
\left[1+a\left(1+z^2 \right) \right]dz,\quad \beta=\sqrt{\frac{a-1}{a}}.
\label{25}
\end{eqnarray}
From (\ref{25}) we obtain the known expression
for the probability per unit time
of photon decay (or creation of the electron-positron pair by the virtual
photon) for $(\hbar k)^2 > 4m^2$ in the medium (see, e.g. \cite{7a},~Sec.113)
\begin{equation}
W \equiv \frac{dw_p}{dt}=\frac{2e^2m^2}{3\hbar \omega}\sqrt{\frac{a-1}{a}}
(2a+1)=\frac{e^2m^2}{3\hbar \omega}
\sqrt{\frac{(\hbar k)^2-4m^2}{(\hbar k)^2}}\left((\hbar k)^2 +2m^2\right).
\label{26}
\end{equation}

4.~We consider now some features of radiation and pair creation
processes in a superdense medium
of the type which exists in white dwarfs. In the such medium the
Landau-Pomeranchuk-Migdal (LPM) effect (suppression of the bremsstrahlung
due to the multiple scattering of a projectile \cite{8})
affects the bremsstrahlung
process for any energy of radiated photon $\hbar \omega$ including
the region where $\hbar \omega \sim \varepsilon$ for the initial
energy $\varepsilon \geq 100$~MeV.
Note that for heavy elements on the Earth this situation
takes place starting from the initial electron energy
of the order of a few TeV. For estimations we use
the results of our papers \cite{9}, \cite{5}.
The function $\nu_0$
is the important characteristics of the LPM effect which defines 
(for $\nu_0 > 1$) the
mean square of the momentum transfer measured in the electron mass $m$
on the formation length of radiation (see Eq.(2.36) of \cite{9}):
\begin{equation}
\nu_0^2=\frac{16\pi nZ^2e^4\varepsilon \varepsilon'}{m^4\omega}
\ln \frac{183 Z^{-1/3}}{\varrho_c},\quad \nu_0^2(\varrho_c)\varrho_c^4=1,
\label{27}
\end{equation}
where $n$ is the number density of atoms of the medium, $Z$
is the atomic number. When $\nu_0 \gg 1$ the standard (Bethe-Heitler)
probability of bremsstrahlung is suppressed due to the LPM effect.
For this case we introduce
\[
\nu_0=\xi_0 \sqrt{\frac{\varepsilon'}{\hbar \omega}},\quad
\xi_0 \simeq \sqrt{\frac{\varepsilon}{\varepsilon_0}}
\left(1+\frac{1}{8\ln \left(183 Z^{-1/3} \right)}
\ln \frac{\varepsilon}{\varepsilon_0} \right),
\]
where $\varepsilon_0=m\left(16\pi Z^2 \alpha^2 \lambda_c^3 n
\ln \left(183 Z^{-1/3} \right) \right)^{-1}$, here $\alpha=e^2/\hbar=1/137$,
$\lambda_c=\hbar/m$. The accurate definition of $\xi_0$ 
follows from (\ref{27}).
For definiteness we consider $Z$=26 (iron), $\rho=10^8~g/cm^3$
and $\varepsilon=200$~MeV, then we have $\varepsilon_0=1.1$MeV,~
$\xi_0 \simeq 15.5$. It is apparent that for such
value of $\nu_0$ the energy losses of a projectile diminishes and
the radiation length increases (see Eq.(2.43) of \cite{9})
\begin{eqnarray}
&& \frac{1}{\varepsilon}\frac{d\epsilon}{dl}=L_{rad}^{-1}
\simeq \frac{\alpha m^2}{2\pi \hbar \varepsilon}
\int_{0}^{\varepsilon}\frac{\varepsilon^2+\varepsilon'^2}
{\varepsilon \varepsilon'} \frac{\nu_0}{\sqrt{2}}\frac{\hbar 
\omega d\hbar \omega}
{\varepsilon^2} \nonumber \\
&& =\frac{\xi_0 \alpha m^2}{2\sqrt{2}\pi \hbar \varepsilon}
\left[B\left(\frac{3}{2}, \frac{1}{2} \right)+
B\left(\frac{3}{2}, \frac{5}{2} \right) \right] =
\frac{9\xi_0 \alpha m^2}{32\sqrt{2} \hbar \varepsilon}
\simeq \frac{3\alpha m^2}{\hbar \varepsilon},
\label{28}
\end{eqnarray}
where $B(x,y)$ is the Euler beta function.
For chosen energy $\varepsilon$ we have $L_{rad} \simeq 7 \cdot 10^{-7}$cm.
At $\hbar \omega \sim \varepsilon$ value of $\nu_0$ increases as
$\sqrt{\gamma}$. The same behavior has $L_{rad} \propto \gamma/\xi_0
\propto \sqrt{\gamma}$.
The formation length in this case is
\begin{equation}
l_f=\frac{2\gamma \hbar}{m(1+\nu_0)}=\frac{\gamma \hbar}{8m} \simeq 2 \cdot
10^{-9} cm.
\label{29}
\end{equation}
Note that under the selected conditions
the value of $\displaystyle{\varphi_0=\frac{(\hbar \omega_0)^2}{m^2}
\frac{\varepsilon \varepsilon'}{(\hbar \omega)^2}}$ is rather small since
$\hbar \omega_0=0.2$~MeV and one can neglect the polarization of the medium.

The differential probability of the pair creation can be found
using the substitution law Eq.(\ref{17}). The lifetime of a photon is
\begin{eqnarray}
&& \frac{1}{\tau}=W_p=\frac{\alpha m^2}{2\pi \hbar^2 \omega}
\int_{0}^{\hbar \omega}\frac{\varepsilon^2+\varepsilon'^2}
{\varepsilon \varepsilon'} \frac{\nu_p}{\sqrt{2}}\frac{d\varepsilon}
{\hbar \omega},
\nonumber \\
&& \nu_p=\xi_p \sqrt{\frac{\varepsilon \varepsilon'}{(\hbar \omega)^2}},\quad
\xi_p=\xi_0 (\varepsilon \rightarrow \hbar \omega),
\label{30}
\end{eqnarray}
here $\varepsilon'=\omega-\varepsilon$.
For the used above parameters and $\omega=200$~MeV we have
\begin{equation}
\frac{1}{\tau}=W_p \simeq \frac{\xi_p \alpha m^2}{\sqrt{2}\pi \hbar^2 \omega}
B\left(\frac{1}{2} , \frac{5}{2} \right)=
\frac{3 \xi_p \alpha m^2}{8\sqrt{2} \hbar^2 \omega} \simeq \frac{4\alpha m^2}
{\hbar^2 \omega},\quad \tau=\frac{3}{4}L_{rad} \simeq 5.3 \cdot 10^{-7}~cm,
\label{31}
\end{equation}
and the formation length of pair creation $l_p$ for $\varepsilon=\omega/2$
is twice shorter than $l_f$ ($l_p \simeq 10^{-9} cm$).

It is shown in \cite{9} that when a projectile crosses boundary
between vacuum and a medium it radiates boundary photons. The transition
radiation can be considered as a particular mechanism of radiation
of boundary photons. We consider the complete probability of the boundary
radiation for $\nu_0 \gg 1$ (see Eq.(4.14) of \cite{9}). Using this formula
we have the contribution of boundary photons in the spectral distribution
of energy losses
\begin{equation}
\frac{d\Delta E_b}{d\hbar \omega}=
\frac{\alpha}{2\pi}\left\{r_1+r_2 \left[\ln \nu_0 -1 -C -
\ln 2+\frac{\sqrt{2}}{\nu_0}\left(\frac{\pi^2}{24}+
\ln \nu_0 +1-C+\frac{\pi}{4} \right) \right]\right\},
\label{32}
\end{equation}
where we put $\kappa=1$~(since $\varphi_0 \ll 1$). We present $\nu_0^2$
Eq.(\ref{27}) as $\nu_0=\xi_0 \sqrt{\varepsilon'/\omega}$. Because we consider
situation when $\xi_0 \gg 1$ we can put $\varepsilon$
the upper limit of the integration over $\omega$ since integrals are
convergent. As a result we find for the energy losses due to boundary photons
radiation
\begin{equation}
\Delta E_b=
\frac{2\alpha \varepsilon}{3\pi}\left[\ln \xi_0 -\frac{9}{16} -C -
\ln 2+\frac{27\pi}{32\sqrt{2} \xi_0}\left(\frac{\pi^2}{24}+
\ln \xi_0 +\frac{4}{27}-C+\frac{\pi}{4} \right) \right].
\label{33}
\end{equation}
For the parameters used ($\xi_0 \simeq 15.5$)
\begin{equation}
\Delta E_b=1.3 \frac{2\alpha \varepsilon}{3\pi} \simeq 2 \cdot 10^{-3}
\varepsilon
\label{34}
\end{equation}

Now we turn to the boundary pair creation. Performing the substitutions
Eq.(\ref{17}) in Eq. (\ref{32}) and integrating over the electron energy
$\varepsilon$ we obtain the total probability of boundary pair creation
\begin{equation}
w_p^b=
\frac{\alpha}{3\pi}\left[\ln \xi_p -\frac{7}{12} -C -
\ln 2+\frac{9\pi}{4\sqrt{2} \xi_p}\left(\frac{\pi^2}{24}+
\ln \xi_p +\frac{5}{6}-C+\frac{\pi}{4}-2\ln 2 \right) \right],
\label{35}
\end{equation}
here we define $\nu_p$ Eq.(\ref{30}) as $\nu_p=\xi_p
\sqrt{\varepsilon \varepsilon'/\omega^2}$. So we have for $\xi_p \simeq 15.5$
\[
w_p^b=1.8\frac{\alpha}{3\pi} \simeq 1.4 \cdot 10^{-3}.
\]
Comparing the energy losses due to the boundary photon radiation (\ref{34})
and due to the transition radiation (\ref{14a}) (for the parameters used
$\hbar \omega_0=0.2~$MeV or
\newline $a=0.04 \ll 1$) we find
\begin{equation}
\frac{\Delta E_b}{\Delta E_{tr}} \simeq \frac{1.3}{\pi \sqrt{a}}
\simeq 2.1,
\label{36}
\end{equation}
so that $\Delta E_b$ is slightly larger than $\Delta E_{tr}$.
For the pair creation we have different situation: the probability of
boundary pair creation $w_p^b$ (\ref{35}) is essentially larger than
the probability of transition pair creation $w_p$ (\ref{22}):
\begin{equation}
\frac{w_p}{w_p^b} \simeq \frac{a^2}{5} \simeq 3 \cdot 10^{-4}.
\label{37}
\end{equation}

Of course, one have to take into account that all the
discussed boundary effects
give visible contribution on the depth of the order of a few formation
lengths.

\newpage

{\bf Figure captions}

\vspace{15mm}
\begin{itemize}

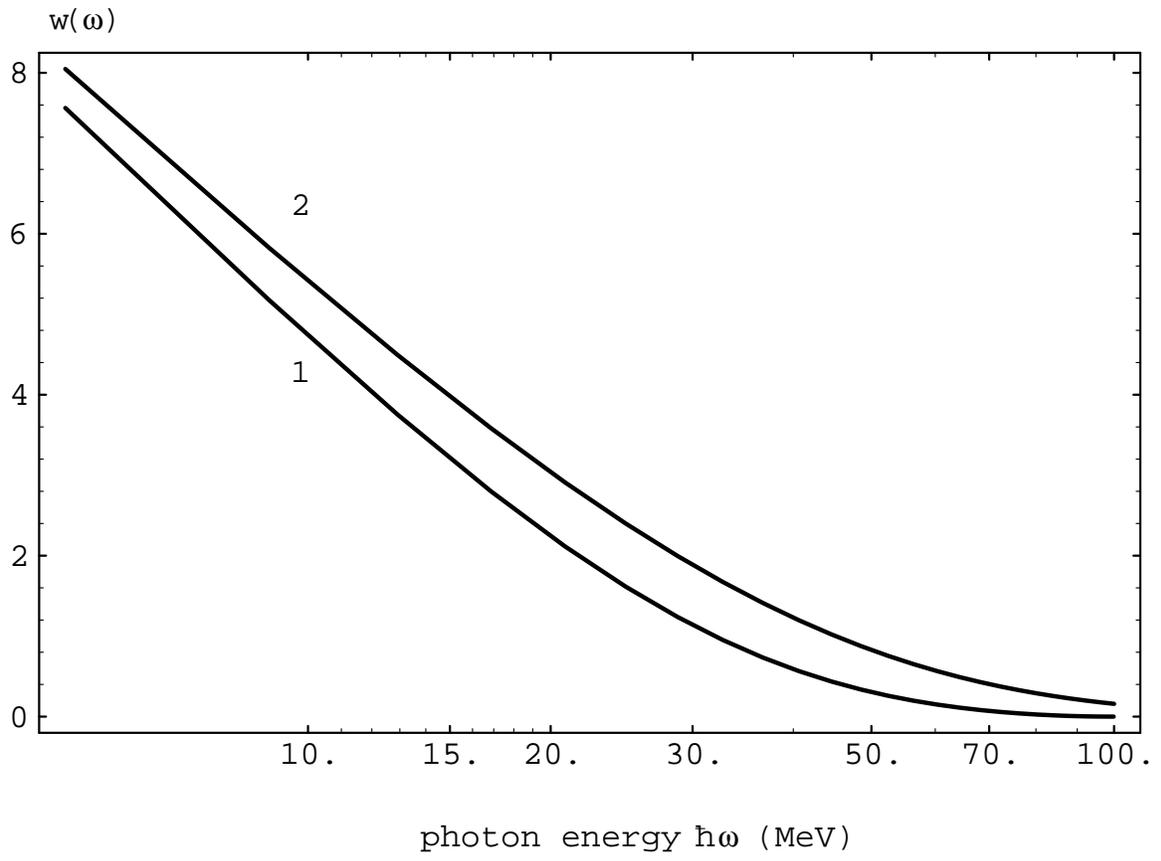
\item {\bf Fig.1} The spectrum $\displaystyle{w(\omega) \equiv
\frac{dw_{tr}}{d\hbar \omega}}$~Eq.(\ref{12})
for the case $\hbar \omega_0=2m,~\varepsilon=100~MeV$
(curve 1) and the classical spectrum (curve 2).

\item {\bf Fig.2} The total energy losses $\Delta E$
when the electron intersects
the boundary between vacuum and a medium versus
$\displaystyle{a=\frac{(\hbar \omega_0)^2}{4m^2}=\frac{(\hbar k)^2}{4m^2}}$.

\item {\bf Fig.3}
The total probability of the transition pair creation $w_p$
as the function of
$\displaystyle{a=\frac{(\hbar \omega_0)^2}{4m^2}=\frac{(\hbar k)^2}{4m^2}}$.

\end{itemize}

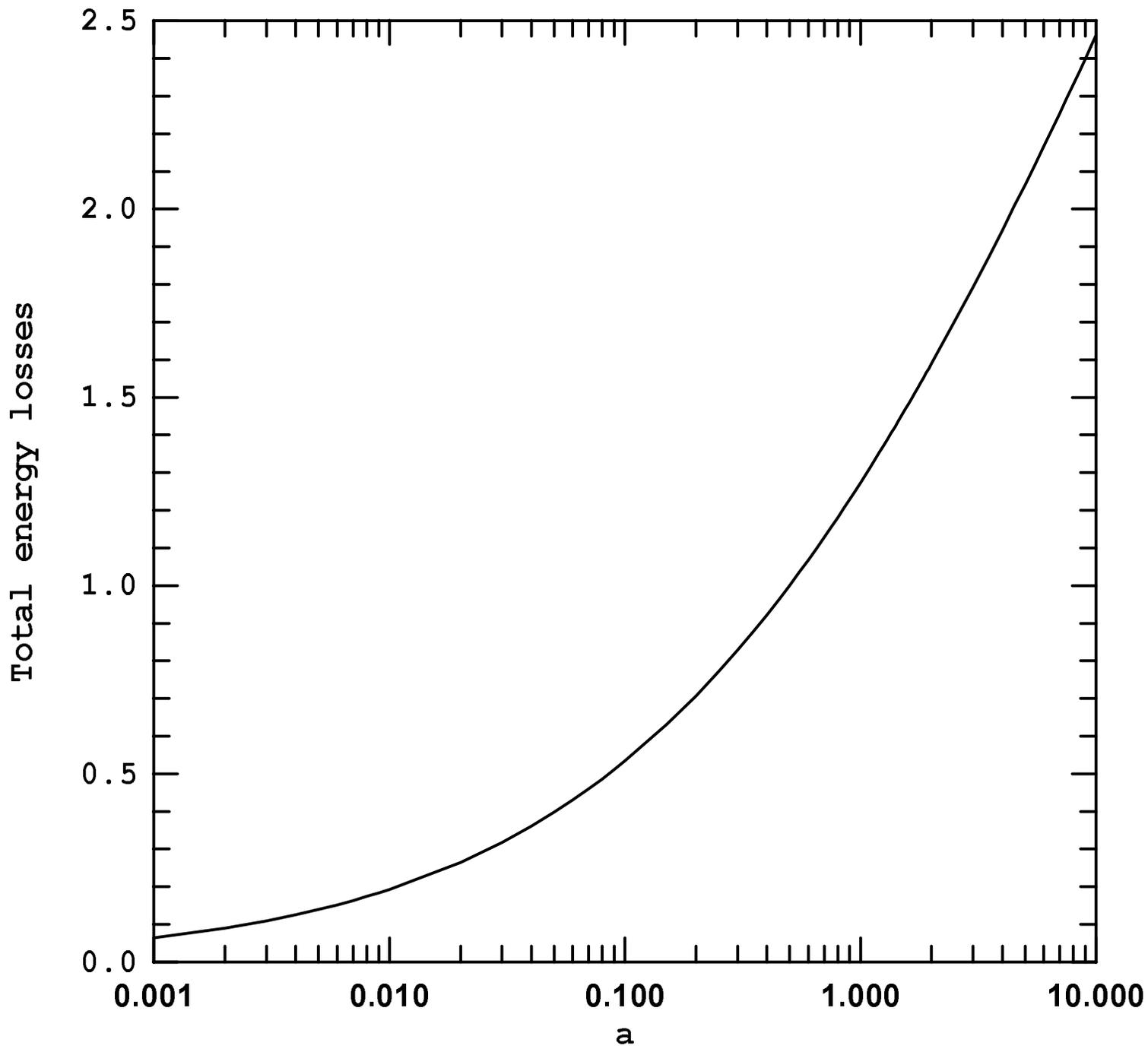
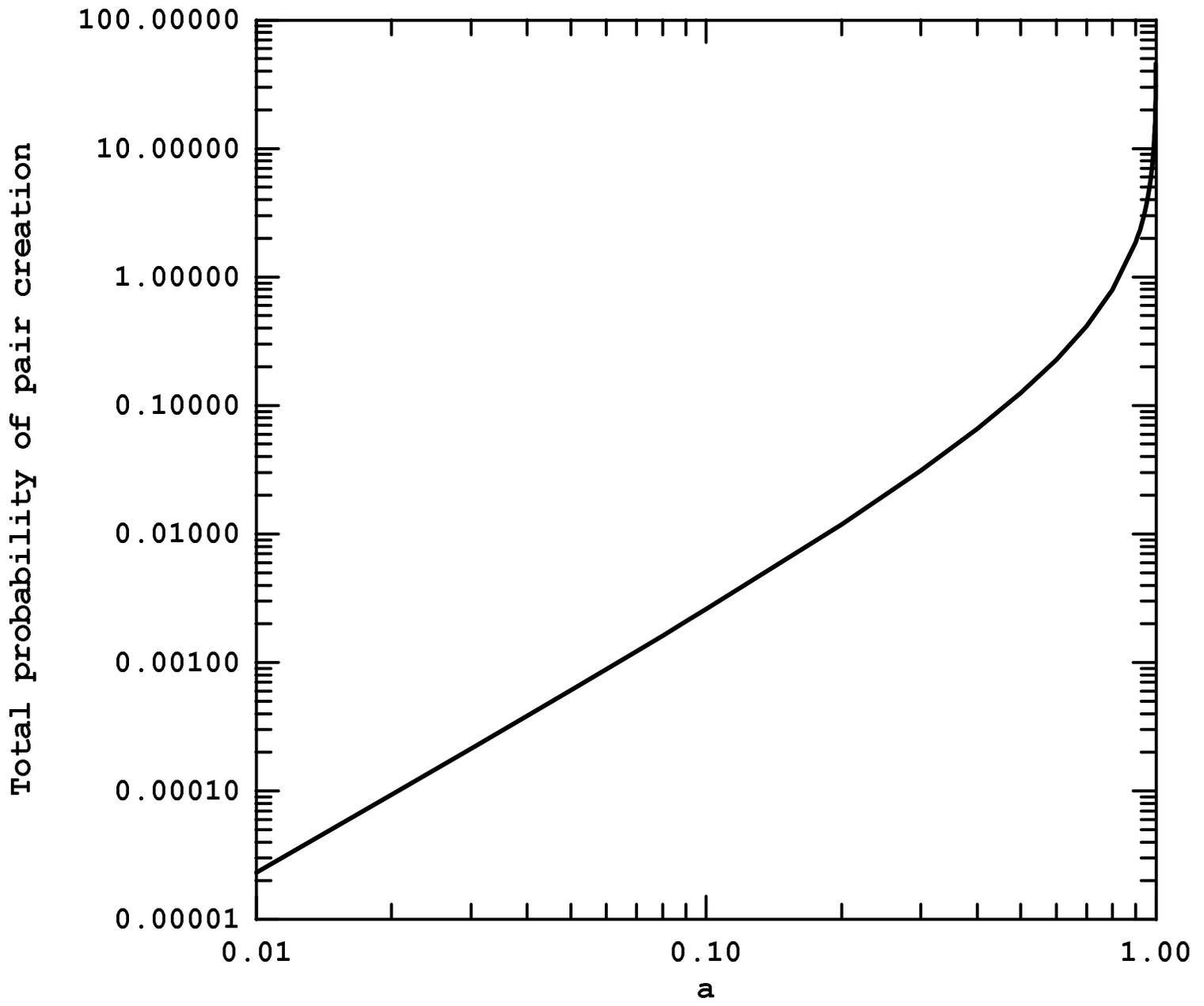
\end{document}